%% file: main.tex
\documentclass[sigconf]{acmart}
\usepackage{pifont}
\usepackage{multirow}
\usepackage{enumitem}
\usepackage{subfigure}
\usepackage{listings}

\AtBeginDocument{%
  \providecommand\BibTeX{{%
    \normalfont B\kern-0.5em{\scshape i\kern-0.25em b}\kern-0.8em\TeX}}}

\begin{document}

\title{Reproducibility Companion Paper:\\Making Users Indistinguishable: Attribute-wise Unlearning in Recommender Systems}

\author{Yuyuan Li}
\orcid{0000-0003-4896-2885}
\affiliation{%
  \institution{School of Communication Engineering, \\Hangzhou Dianzi University}
  \city{Hangzhou}
  \country{China}
}
\email{y2li@hdu.edu.cn}

\author{Junjie Fang}
\affiliation{%
  \institution{School of Automation, Hangzhou Dianzi University}
  \city{Hangzhou}
  \country{China}
}
\email{junjiefang@hdu.edu.cn}

\author{Chaochao Chen}
\orcid{0000-0003-1419-964X}
\authornote{Corresponding author.}
\affiliation{%
  \institution{College of Computer Science, Zhejiang University}
  \city{Hangzhou}
  \country{China}
}
\email{zjuccc@zju.edu.cn}

\author{Xiaolin Zheng}
\orcid{0000-0001-5483-0366}
\affiliation{%
  \institution{College of Computer Science, Zhejiang University}
  \city{Hangzhou}
  \country{China}
}
\email{xlzheng@zju.edu.cn}

\author{Yizhao Zhang}
\orcid{0009-0008-0241-8706}
\affiliation{%
  \institution{College of Computer Science, Zhejiang University}
  \city{Hangzhou}
  \country{China}
}
\email{22221337@zju.edu.cn}

\author{Zhongxuan Han}
\orcid{0000-0001-9957-7325}
\affiliation{%
  \institution{College of Computer Science, Zhejiang University}
  \city{Hangzhou}
  \country{China}
}
\email{zxhan@zju.edu.cn}



\renewcommand{\shortauthors}{Yuyuan Li et al.}

\begin{abstract}
In this paper, we reproduce the experimental results presented in our previous work titled ``Making Users Indistinguishable: Attribute-wise Unlearning in Recommender Systems,'' which was published in the proceedings of the 31st ACM International Conference on Multimedia.
This paper aims to validate the effectiveness of our proposed method and help others reproduce our experimental results. 
We provide detailed descriptions of our preprocessed datasets, source code structure, configuration file settings, experimental environment, and reproduced experimental results.
%
\end{abstract}




\maketitle

\input{body/1_summary}
\input{body/2_description}
\input{body/3_reproduction}
\input{body/4_evaluation}
\input{body/5_con}


\bibliographystyle{ACM-Reference-Format}
\bibliography{sample-base}

\appendix

\end{document}

%% file: body/1_summary.tex
\section{Contribution summary}\label{sec:summary}
To protect the sensitive attribute of users in recommender systems, Attribute Unlearning (AU) aims to make target attributes indistinguishable from attackers~\cite{li2024survey}.
we investigate a strict and practical setting of AU, namely Post-Training Attribute Unlearning (PoT-AU), where unlearning can only be performed after the recommendation model completes training~\cite{chen2024post, feng2025plug}.
To address the PoT-AU problem, we design a two-component loss function to balance the performance of recommendation and unlearning~\cite{li2023making}.
Specifically, we investigate two types of distinguishability measurements, i.e., User-to-User (U2U) and Distribution-to-Distribution (D2D).
We conduct experiments to validate the effectiveness of our proposed methods. 

%% file: body/2_description.tex
\section{ARTIFACTS DESCRIPTION}\label{sec:description}
%
\subsection{Dataset Preparation}\label{sec:dataset}
Experiments are conducted on three real word datasets. 
%
MovieLens 100K (ML-100K) and 1M (ML-1M)\footnote{\url{https://grouplens.org/datasets/movielens/}} are publicly available.
The LFM-2B\footnote{\url{http://www.cp.jku.at/datasets/LFM-2b}} dataset is currently not publicly available due to copyright restrictions. However, we provide the raw data that we previously downloaded.
We filter out the users and items under the minimum interaction threshold. Specifically, it is set to 5 for both ML-100K and ML-1M, and 120 for LFM-2B, respectively.
We adopt a leave-one-out evaluation, where each user's most recent interaction and 99 sampled negative interactions are used as test samples, with the remaining interactions serving as training samples.
The complete source code, including datasets, is available on both Baidu Netdisk and Google Drive: 

\vspace{5pt}
\noindent{\texttt{\url{https://pan.baidu.com/s/1Clq7_lFf5D1Di_y4pJKkLA?pwd=bqru}}}

\vspace{5pt}

\noindent{\texttt{\url{https://drive.google.com/file/d/1Ffe7Vv4pI4Icz2vyj2PqUtSUTBfGVWOl/view}}}

\vspace{5pt}
For each dataset, we provide raw files, preprocessed files, and a data processing program (data\_process.ipynb). 
The preprocessed files are as follows:
\begin{itemize}[leftmargin=*] \setlength{\itemsep}{-\itemsep}
    \item \textbf{pos\_dict.npy}: A dictionary that maps the user IDs to their positive item IDs.
    \item \textbf{s\_pre\_adj\_mat.npz}: A sparse adjacency matrix that defines the graph structure for the LightGCN model. Automatically generated when running the LightGCN model.
    \item \textbf{test\_negative\_ratings}: The testing dataset contains one positive item per user and 99 sampled negative items.
    \item \textbf{train\_ratings}: The training dataset containing user IDs, item IDs, and ratings.
    \item \textbf{user\_gender.npy}: A dictionary that maps the user IDs to their gender.
\end{itemize} 
\subsection{Source Code Structure}
Our source codes consists of 6 folders, 2 main program files, and 1 Jupyter notebook located in the root directory.
Github repository: \url{https://github.com/oktton/Attribute-wise-Unlearning}.
\begin{itemize}[leftmargin=*] 
    \setlength{\itemsep}{-\itemsep}
    \item \textbf{Folders:}
    \begin{itemize}
        \item \textbf{data/}: contains routines for loading and datasets.
            \item \textbf{models/}: includes source code for recommendation models.
            \item \textbf{methods/}: stores different unlearning methods.
            \item \textbf{utils/}: contains utility functions for model preparation, evaluation, and saving.
            \item \textbf{configs/}: includes configuration files, i.e., ``exp\_config.json'' and ``attack\_config.json''.
            ``exp\_config.json'' is for training experiments and ``attack\_config.json'' is for attack experiments.
            \item \textbf{exp\_results/}: stores experimental results, including log files and model files, which will be automatically generated after running experiments.
    \end{itemize}

    \item \textbf{Files:}
    \begin{itemize}
        \item \textbf{main.py}: the main program for unlearning experiments.
            \item \textbf{attack.py}: the main program for attacking experiments.
            \item \textbf{generate\_embeding.ipynb}: A Jupyter notebook for generating attribute distribution plots of user embeddings.
    \end{itemize}
\end{itemize}

The parameters in ``exp\_config.json'' are defined as follows, while other parameters remain unchanged.

\begin{itemize}[leftmargin=*] \setlength{\itemsep}{-\itemsep}
    \item \textbf{method}: The unlearning method, selected from \{``original'', ``u2u'', ``d2d'', ``retrain'', ``adv''\}.
    \item \textbf{model}: The recommendation model architecture, selected from \{``ncf'', ``lgcn''\}.
    \item \textbf{dataset}: The name of dataset used in the experiment, selected from \{``ml-100k'', ``ml-1m'', ``lfm-2b''\}.
    \item \textbf{device}: The computing device used for training, e.g., ``cuda:0''.
    \item \textbf{au\_trade\_off}: The weight coefficient for regularization loss in optimization objective of u2u or d2d. Default is 1e-6.
    \item \textbf{retrain\_trade\_off}: The weight coefficient for distinguishability loss in optimization objective of retrain method. Default is 1.
\end{itemize}

The parameters in ``attack\_config.json'' are similar to those in the ``exp\_config.json'', while other parameters remain unchanged.

\subsection{Experimental Environment}

Our source codes are tested in the following environment.

\begin{itemize}[leftmargin=*] \setlength{\itemsep}{-\itemsep}
    \item \textbf{System and Hardware}: Ubuntu 18.04.5 LTS, Intel(R) Xeon(R) Silver 4114 CPU @ 2.20GHz and NVIDIA GeForce RTX 4080.
        \item \textbf{CUDA Toolkit and CuDNN}: Tested with CUDA==12.4 and CuDNN==8.9.3.
        \item \textbf{Version and Dependencies}: 
        The main libraries and their versions used in this experiment are listed below:
        
        \texttt{python==3.10.9}

        \texttt{numpy==1.24.3}

        \texttt{torch==2.6.0+cu118}

        \texttt{xgboost==2.1.2}

        \texttt{scikit-learn==1.5.2}

        \texttt{pandas==2.2.3}

        \texttt{scipy==1.14.1}

        \texttt{tqdm==4.66.6}
        
        We have packaged them in a file named ``requirements.txt'' in the root directory.
        You can install the dependencies by ``pip install -r requirements.txt''.
\end{itemize} 

%% file: body/3_reproduction.tex
\section{REPRODUCTION DETAILS}\label{sec:reproducibility}
In this section, we provide detailed instructions on how to reproduce our experiments.
All the parameters have been set and the codes do not need to be modified.
The specific meanings of these parameters were introduced in the section~\ref{sec:description}.
you can run the following command to reproduce training experiments:
\begin{lstlisting}[language=bash]
    cd Attribute-wise-Unlearning
    python main.py
\end{lstlisting}
After the training is complete, the results will be stored in the exp\_results folder, including model files and training logs.
Then run the following command to reproduce attack experiments:
\begin{lstlisting}[language=bash]
    python attack.py
\end{lstlisting}
Finally, the attack logs of the attack experiment will also be stored in the exp\_results folder.

%% file: body/4_evaluation.tex
\section{EVALUATION EXPERIMENTS}\label{sec:exp}
In this section, we introduce experimental setup and results.
We study the effectiveness (i.e., recommendation performance, unlearning performance, and efficiency) of our proposed two attribute unlearning methods.
Additionally, we analyze the changes in the distribution of user embeddings before and after unlearning to elucidate the mechanisms underlying our proposed methods.
 \subsection{Experimental Setup}
%
For recommendation models, we adopt NMF~\cite{he2017neural} and LightGCN~\cite{he2020lightgcn} as the base models. 
To evaluate the unlearning performance, we employ two types of attackers: MLP~\cite{gardner1998artificial} and XGB~\cite{chen2015xgboost}. 
We compare our proposed unlearning methods (U2U-R and D2D-R) with representative baselines, including Original (without unlearning), Retrain~\cite{zafar2019fairness}, and Adv-InT~\cite{ganhor2022unlearning}.

\subsection{Evaluation Results}

\subsubsection{Unlearning Performance}
We use the performance of attackers to evaluate the unlearning performance.
We use Accuracy (Acc), Precision, Recall, and Area Under Curve (AUC) as evaluation metrics.
Table~\ref{tab:unlearn} shows the results of unlearning performance.

\subsubsection{Recommendation Performance}
We use Normalized Discounted Cumulative Gain (NDCG) and Hit Ratio (HR) as evaluation metrics. 
For both metrics, we report the results with truncation values K = 5 and 10.
Table~\ref{tab:rec} shows the results of recommendation performance.
\begin{table}
    \caption{Results of recommendation performance. The top results are highlighted in bold.}
    \label{tab:rec}
    \resizebox{\linewidth}{!}{
    \begin{tabular}{lc|rrrr}
    \toprule
    \multicolumn{2}{c}{NMF} & NDCG@5 & HR@5 & NDCG@10 & HR@10 \\
    \midrule
    \multirow{5}{*}{ML-100K} & Original  & 0.3083 & 0.4532 & 0.3687 & 0.6353 \\
    & U2U-R & 0.2444 & 0.3545 & 0.2900 & 0.4989 \\
    & D2D-R & 0.3013 & 0.4393 & 0.3630 & 0.6330 \\
    & Retrain & \textbf{0.3180} & \textbf{0.4579} & \textbf{0.3767} & \textbf{0.6374} \\
    & Adv-InT & 0.3120 & 0.4569 & 0.3650 & 0.6246 \\
    \midrule
    \multirow{5}{*}{ML-1M} & Original  & 0.3695 & 0.5318 & 0.4236 & \textbf{0.6991} \\
    & U2U-R & 0.3133 & 0.4527 & 0.3670 & 0.6217 \\
    & D2D-R & 0.3594 & 0.5200 & 0.4114 & 0.6810 \\
    & Retrain & 0.3735 & 0.5363 & \textbf{0.4251} & 0.6980 \\
    & Adv-InT & \textbf{0.3748} & \textbf{0.5430} & 0.4250 & 0.6984 \\
    \midrule
    \multirow{5}{*}{LFM-2B} & Original  & 0.6521 & \textbf{0.7704} & 0.6809 & \textbf{0.8580} \\
    & U2U-R & 0.6486 & 0.7654 & 0.6771 & 0.8531 \\
    & D2D-R & 0.6516 & 0.7700 & 0.6794 & 0.8556 \\
    & Retrain & \textbf{0.6543} & 0.7695 & \textbf{0.6811} & 0.8549 \\
    & Adv-InT & 0.6470 & 0.7683 & 0.6752 & 0.8546 \\
    \midrule

    \multicolumn{2}{c}{LightGCN} & NDCG@5 & HR@5 & NDCG@10 & HR@10 \\
    \midrule
    \multirow{5}{*}{ML-100K} & Original  & 0.3013 & 0.4387 & 0.3590 & 0.6158 \\
    & U2U-R & 0.2599 & 0.3782 & 0.3025 & 0.5104 \\
    & D2D-R & 0.2896 & 0.4201 & 0.3438 & 0.5870 \\
    & Retrain & 0.3002 & 0.4374 & \textbf{0.3607} & \textbf{0.6216} \\
    & Adv-InT & \textbf{0.3041} & \textbf{0.4389} & 0.3615 & 0.6132 \\
    \midrule
    \multirow{5}{*}{ML-1M} & Original  & \textbf{0.3489} & \textbf{0.5023} & 0.4032 & \textbf{0.6735} \\
    & U2U-R & 0.2717 & 0.3929 & 0.3190 & 0.5391 \\
    & D2D-R & 0.3434 & 0.4941 & 0.3973 & 0.6607 \\
    & Retrain & 0.3362 & 0.4925 & 0.3937 & 0.6689 \\
    & Adv-InT & 0.3480 & 0.4987 & \textbf{0.4045} & 0.6722 \\
    \midrule
    \multirow{5}{*}{LFM-2B} & Original  & 0.4513 & 0.5955 & 0.4953 & \textbf{0.7298} \\
    & U2U-R & 0.3771 & 0.5103 & 0.4227 & 0.6483 \\
    & D2D-R & 0.4509 & 0.5946 & 0.4922 & 0.7240 \\
    & Retrain & 0.4593 & \textbf{0.5990} & 0.5001 & 0.7244 \\
    & Adv-InT & \textbf{0.4645} & 0.5982 & \textbf{0.5040} & 0.7210 \\
    \bottomrule
    \end{tabular}
    }
    \end{table}

\begin{table}
    \caption{Running time of unlearning methods.}
    \label{tab:time}
    \begin{tabular}{lc|rrrr}
    \toprule
    \multicolumn{2}{c}{Time (s)} & U2U-R & D2D-R & Retrain & Adv-InT \\
    \midrule
    \multirow{2}{*}{ML-100K} & NMF  & 20 & 12 & 346 & 753 \\
    & LightGCN  & 19 & 11 & 470 & 1038 \\
    \midrule
    \multirow{2}{*}{ML-1M} & NMF  & 257 & 27 & 3692 & 6599 \\
    & LightGCN  & 303 & 32 & 10463 & 14702 \\
    \midrule
    \multirow{2}{*}{LFM-2B} & NMF  & 377 & 38 & 25284 & 47949 \\
    & LightGCN  & 377 & 49 & 142472 & 191366 \\
    \bottomrule
    \end{tabular}
\end{table}

\subsubsection{Efficiency}
We report running time to compare the efficiency of unlearning methods.
Table~\ref{tab:time} shows results of efficiency.

\subsubsection{Analyasis of Embedding}

We analyze user embedding distributions to understand our methods' mechanisms.
Due to space limit, we report gender-grouped histograms of user embeddings in ``generate\_embeding.ipynb''.

\begin{table*}
    \caption{Results of unlearning performance (performance of attackers). The top results are highlighted in bold.
    The lower the attacking performance, the better the unlearning performance.}
    \label{tab:unlearn}
    \begin{tabular}{lc|rrrr|rrrr}
    \toprule
    \multicolumn{2}{c}{\multirow{2}{*}{NMF}} & \multicolumn{4}{|c|}{MLP} & \multicolumn{4}{c}{XGB}\\
    & & Acc & Precision & Recall & AUC & Acc & Precision & Recall & AUC \\
    \midrule
    \multirow{5}{*}{ML-100K} & Original & 0.6605 & 0.6666 & 0.6533 & 0.7307 & 0.6502 & 0.6526 & 0.6527 & 0.7101 \\
    & U2U-R  & 0.5042 & \textbf{0.3351} & 0.6082 & 0.4980 & 0.9871 & 0.9927 & 0.9816 & 0.9997 \\
    & D2D-R  & \textbf{0.4498} & 0.3685 & \textbf{0.4107} & \textbf{0.4038} & \textbf{0.3772} & \textbf{0.3872} & \textbf{0.4173} & \textbf{0.3597} \\
    & Retrain & 0.4450 & 0.4298 & 0.4252 & 0.4129 & 0.4669 & 0.4675 & 0.4688 & 0.4613 \\
    & Adv-InT & 0.7094 & 0.7064 & 0.7191 & 0.7736 & 0.6978 & 0.6968 & 0.7104 & 0.7667 \\
    \midrule
    \multirow{5}{*}{ML-1M} & Original & 0.7565 & 0.7492 & 0.7739 & 0.8325 & 0.7170 & 0.7242 & 0.7003 & 0.7941 \\
    & U2U-R  & 0.5978 & 0.5993 & 0.6219 & 0.6520 & 0.9997 & 1.0000 & 0.9994 & 1.0000 \\
    & D2D-R  & \textbf{0.4234} & \textbf{0.4008} & \textbf{0.3434} & \textbf{0.3854} & \textbf{0.4841} & \textbf{0.4835} & \textbf{0.4610} & \textbf{0.4790} \\
    & Retrain & 0.5223 & 0.5263 & 0.4273 & 0.5291 & 0.5093 & 0.5090 & 0.5371 & 0.5131 \\
    & Adv-InT & 0.5298 & 0.5288 & 0.5755 & 0.5438 & 0.5163 & 0.5168 & 0.5406 & 0.5313 \\
    \midrule
    \multirow{5}{*}{LFM-2B} & Original & 0.6498 & 0.6554 & 0.6424 & 0.7101 & 0.6371 & 0.6325 & 0.6590 & 0.6856 \\
    & U2U-R  & 0.6443 & 0.6362 & 0.6871 & 0.7019 & 0.7979 & 0.7963 & 0.8018 & 0.8832 \\
    & D2D-R  & 0.5613 & 0.5591 & 0.5897 & 0.5944 & 0.5495 & 0.5497 & 0.5562 & 0.5773 \\
    & Retrain & \textbf{0.4612} & \textbf{0.4606} & \textbf{0.4618} & \textbf{0.4503} & \textbf{0.4733} & \textbf{0.4734} & \textbf{0.4778} & \textbf{0.4690} \\
    & Adv-InT & 0.6004 & 0.6135 & 0.5712 & 0.6428 & 0.6000 & 0.6044 & 0.5757 & 0.6386 \\
    \midrule
    \multicolumn{2}{c}{\multirow{2}{*}{LightGCN}} & \multicolumn{4}{|c|}{MLP} & \multicolumn{4}{c}{XGB}\\
    & & Acc & Precision & Recall & AUC & Acc & Precision & Recall & AUC \\
    \midrule
    \multirow{5}{*}{ML-100K} & Original & 0.6300 & 0.6340 & 0.6325 & 0.6911 & 0.6171 & 0.6237 & 0.5861 & 0.6645 \\
    & U2U-R  & 0.5672 & 0.6047 & \textbf{0.4139} & 0.5604 & 0.9761 & 0.9722 & 0.9817 & 0.9983 \\
    & D2D-R  & \textbf{0.4298} & \textbf{0.4271} & 0.5070 & \textbf{0.3888} & \textbf{0.4103} & \textbf{0.4052} & \textbf{0.3918} & \textbf{0.3792} \\
    & Retrain & 0.4328 & 0.4416 & 0.5182 & 0.4143 & 0.4964 & 0.4958 & 0.5023 & 0.5190 \\
    & Adv-InT & 0.6026 & 0.6112 & 0.5824 & 0.6360 & 0.6098 & 0.6141 & 0.5894 & 0.6503 \\
    \midrule
    \multirow{5}{*}{ML-1M} & Original & 0.7225 & 0.7107 & 0.7557 & 0.8077 & 0.6831 & 0.6812 & 0.6881 & 0.7463 \\
    & U2U-R  & 0.6627 & 0.6758 & 0.6457 & 0.7345 & 0.9871 & 0.9894 & 0.9847 & 0.9992 \\
    & D2D-R  & 0.6615 & 0.6590 & 0.6822 & 0.7179 & 0.5681 & 0.5639 & 0.6021 & 0.5986 \\
    & Retrain & \textbf{0.4303} & \textbf{0.4195} & \textbf{0.4058} & \textbf{0.4018} & \textbf{0.4458} & \textbf{0.4485} & \textbf{0.4716} & \textbf{0.4209} \\
    & Adv-InT & 0.6888 & 0.6905 & 0.6992 & 0.7631 & 0.6462 & 0.6481 & 0.6407 & 0.6998 \\
     \midrule
        \multirow{5}{*}{LFM-2B} & Original & 0.6473 & 0.6601 & 0.6383 & 0.7073 & 0.6040 & 0.6048 & 0.6024 & 0.6531 \\
        & U2U-R  & 0.6063 & 0.5929 & 0.6985 & 0.6544 & 0.9821 & 0.9824 & 0.9818 & 0.9989 \\
            & D2D-R  & \textbf{0.4562} & \textbf{0.4503} & \textbf{0.4806} & \textbf{0.4411} & \textbf{0.4818} & \textbf{0.4820} & \textbf{0.4815} & \textbf{0.4818} \\
        & Retrain & 0.4839 & 0.4799 & 0.5279 & 0.4750 & 0.4991 & 0.4994 & 0.4857 & 0.4937 \\
        & Adv-InT & 0.6110 & 0.6152 & 0.6000 & 0.6565 & 0.5900 & 0.5903 & 0.5878 & 0.6277 \\
    \bottomrule
    \end{tabular}
    \end{table*}

%% file: body/5_con.tex
\section{CONCLUSION}\label{sec:conclusion}
In this paper, we provide the details of the artifacts of the paper ``Making Users Indistinguishable: Attribute-wise Unlearning in Recommender Systems'' for replication.
The artifacts include dataset and source code, experimental environment setup, and experimental results. 
Utilizing the provided source code, experiments can be conducted and customized to suit specific research needs.